\documentclass[conference]{IEEEtran}

\usepackage[T1]{fontenc}
\usepackage[utf8]{inputenc}
\usepackage{cite}
\usepackage{url}
\usepackage{booktabs}
\usepackage{array}
\usepackage{tabularx}
\usepackage{enumitem}
\usepackage[hidelinks]{hyperref}

\newcommand{\framework}{Agentic Agile-V}
\newcommand{\scopev}{SCOPE-V}

\begin{document}

\title{Agentic Agile-V: From Vibe Coding to Verified Engineering in Software and Hardware Development}

\author{\IEEEauthorblockN{Christopher Koch}
\IEEEauthorblockA{Independent Researcher}}

\maketitle

\begin{abstract}
Agentic AI coding systems can inspect repositories, plan implementation steps, edit files, call tools, run tests, and submit pull requests. These capabilities make software and hardware development faster in some settings, but current evidence does not support the simple claim that autonomous code generation automatically improves engineering outcomes. Controlled studies report productivity gains in some enterprise tasks, slowdowns in mature open-source work, moderate but heterogeneous meta-analytic effects, and persistent failures in repository setup, dependency handling, permission gating, and hardware verification. This paper argues that the central problem is no longer prompt engineering; it is engineering process control. It synthesizes evidence from agentic software engineering, GitHub-scale adoption studies, repository-level agent configuration, productivity trials, issue-resolution benchmarks, and hardware/RTL verification research. It proposes \framework{}, a process framework that uses Agile-V as the lifecycle backbone and a task-level \scopev{} loop - Specify, Constrain, Orchestrate, Prove, Evolve, and Verify - to convert conversational intent into structured engineering artifacts and acceptance evidence. The paper contributes: (i) a taxonomy of minimum input artifacts for agentic software, firmware, and hardware work; (ii) a conversation-to-contract gate that separates exploratory dialogue from implementation; (iii) risk-adaptive feature, bug-fix, testing, and hardware workflows; and (iv) an evidence-bundle acceptance model for agent-generated artifacts. The paper concludes that agentic AI does not eliminate engineering discipline; it increases the value of requirements, constraints, traceability, independent verification, and human approval.
\end{abstract}


\section{Introduction}
Agentic AI changes the form of engineering work. Modern coding agents can do more than complete a line of code: they can read repositories, plan work, invoke terminals, modify multiple files, run tests, prepare pull requests, and respond to review feedback. The ecosystem now includes asynchronous coding agents, repository-level instruction files, agent SDKs, multi-agent workspaces, and model-agnostic routing. Large-scale evidence shows that this is already an operational phenomenon rather than a speculative one: the AIDev dataset reports 932,791 agent-authored pull requests across 116,211 repositories and 72,189 developers \cite{aidev}. Surveys on LLM-based agents for software engineering identify applications across requirements, implementation, testing, maintenance, and human-agent collaboration \cite{agent4se,codeagent_survey}.

Yet the evidence base also challenges the strongest hype. A randomized controlled trial with 96 full-time Google engineers reported an estimated 21 percent reduction in time on a complex enterprise task with AI assistance \cite{google_rct}. In contrast, a METR randomized trial with experienced open-source developers found that AI tools increased task completion time by 19 percent in mature repositories, despite developer expectations of speedup \cite{metr_rct}. A 2026 meta-analysis found a statistically significant but moderate productivity effect, with substantial heterogeneity and smaller effects in open-source and enterprise contexts \cite{meta_programming}. Repository-level configuration files can reduce runtime and token use in some settings \cite{agentsmd_efficiency}, but other evaluations find that context files can reduce task success and increase cost when they add unnecessary or mismatched requirements \cite{agentsmd_eval}. Hardware evidence is even more cautionary: RealBench reports low pass rates on real-world Verilog generation, including 0 percent pass@1 on system-level tasks for evaluated models \cite{realbench}, and FIXME argues that functional verification remains underexplored despite rapid progress in LLM-aided design \cite{fixme}.

These findings point to a process gap. Agentic AI can generate plausible engineering artifacts faster than humans can inspect them. Therefore, the bottleneck shifts from code synthesis to specification quality, execution context, verification, traceability, and controlled iteration. The right question is not merely ``How do we prompt better?'' but ``Which process turns natural-language intent into verifiable engineering output?''

This paper proposes \framework{}, a process framework for agentic software, firmware, and hardware development. It integrates Agile-V, a compliance-ready approach that combines Agile iteration with V-model verification and audit artifact generation \cite{agilev}, with a task-level \scopev{} execution loop. The central principle is:

\begin{quote}
Conversation is useful for discovering intent; structured artifacts are required for implementation; evidence is required for acceptance.
\end{quote}

The contributions are fourfold:
\begin{enumerate}[leftmargin=*]
    \item A synthesis of evidence from agentic coding research, GitHub adoption studies, repository-configuration work, productivity studies, issue-resolution benchmarks, and hardware verification benchmarks.
    \item A conversation-to-contract model that separates exploratory dialogue from structured execution.
    \item The \framework{} process model, combining Agile-V lifecycle structure with the \scopev{} task loop.
    \item Practical workflows and risk-adaptive evidence gates for feature work, bug fixing, test generation, firmware, and hardware/RTL development.
\end{enumerate}

\section{Background and Evidence Base}
\subsection{From Assistants to Agents}
LLM-based software agents extend standalone language models with perception, planning, memory, tool use, execution environments, and human interaction \cite{agent4se}. Code-generation agents differ from earlier code generators because they can decompose tasks, navigate repositories, execute tests, debug failures, and integrate changes across the software development lifecycle \cite{codeagent_survey}. OpenHands, for example, emphasizes sandboxed execution, lifecycle control, model-agnostic routing, custom tools, memory management, and workspace/API integrations \cite{openhands_sdk}. GitHub has also moved toward agentic workflows in which tasks can be delegated to agents that clone repositories, work in virtual environments, document their decisions, run tests, and propose pull requests for human review \cite{github_agent_news,github_agent_hq}.

This transition changes the engineering problem. The unit of interaction is no longer just a prompt and an answer. It is a socio-technical loop consisting of requirements, repository context, tools, permissions, tests, build environments, review practices, and release gates.

\subsection{Productivity Is Context-Dependent}
The current productivity literature supports a balanced claim. AI assistance can help, especially in well-scoped or routine tasks, but it can also create overhead through prompting, waiting, review, and correction. The Google RCT provides evidence for acceleration in an enterprise setting \cite{google_rct}; the METR RCT shows slowdown in familiar mature open-source work \cite{metr_rct}; and a meta-analysis reports a moderate average effect with strong context dependence \cite{meta_programming}. A systematic literature review also warns that developer productivity is multi-dimensional and cannot be reduced to output volume or task time alone \cite{slr_productivity}. Maintenance-burden research further suggests that AI-assisted output may shift review and rework load toward experienced developers \cite{maintenance_burden}.

The implication is not that agentic AI fails. It is that AI-assisted engineering is a process-sensitive intervention. Task type, codebase complexity, developer expertise, test coverage, dependency setup, and verification cost can determine whether agents help or hurt.

\subsection{GitHub Evidence and Repository Configuration}
GitHub evidence shows rapid adoption. AIDev provides a large corpus of agent-authored pull requests \cite{aidev}. A task-stratified analysis of 7,156 pull requests found that task type strongly influences acceptance rates: documentation changes were accepted more often than new features, and no single agent performed best across all categories \cite{agent_pr_compare}. An empirical study of Claude Code pull requests found high acceptance, but also substantial human revision, especially for bug fixes, documentation, and project-specific standards \cite{claude_prs}.

Configuration artifacts are emerging as the process layer around agents. A study of 2,926 GitHub repositories found eight configuration mechanisms across tools such as Claude Code, GitHub Copilot, Cursor, Gemini, and Codex, with context files dominating and \texttt{AGENTS.md} emerging as an interoperable standard \cite{agent_config}. One study found that \texttt{AGENTS.md} was associated with lower runtime and lower output-token consumption \cite{agentsmd_efficiency}. Another found that context files can reduce task success and increase inference cost when they impose unnecessary requirements \cite{agentsmd_eval}. A factorial study of configuration-file structure found limited evidence that size, position, architecture, or local contradiction variables alone create reliable adherence effects \cite{instruction_adherence}.

These mixed results motivate a minimal-context principle: repository instructions should be short, current, non-contradictory, and tied to executable feedback.

\subsection{Repository Execution Remains Hard}
Real-world repository tasks remain challenging. GitTaskBench found that even the best evaluated system, OpenHands plus Claude 3.7, solved 48.15 percent of tasks, with many failures caused by environment setup and dependency resolution \cite{gittaskbench}. RepoMaster improves repository exploration by constructing graphs and pruning context, showing that repository understanding and context selection are central bottlenecks \cite{repo_master}. SWE-Skills-Bench found that procedural skill packages usually provide limited marginal benefit unless they match the domain and project context \cite{swe_skills}. SWE-rebench V2 further demonstrates the importance of reproducible execution environments and reliable test suites at scale \cite{swe_rebench}.

The lesson is that code context alone is insufficient. Agents need execution context: build commands, dependency setup, test commands, environment variables, toolchain information, simulator access, and clear acceptance criteria.

\subsection{Hardware and Firmware Raise the Bar}
Hardware, firmware, and embedded development have stricter failure modes. Incorrect pin mappings, register values, timing assumptions, bus behavior, reset handling, or memory layout can produce failures that are costly or unsafe. RealBench explicitly addresses the gap between simple Verilog benchmarks and real-world IP-level workflows; its low pass rates show that current LLMs are not reliable system-level hardware generators \cite{realbench}. FIXME focuses on design verification and uses silicon-proven designs to evaluate functional verification capabilities \cite{fixme}. Surveys of LLMs for electronic design automation and hardware/software co-design highlight both opportunities and reliability limitations \cite{eda_survey,hw_sw_survey,hardware_limits}.

For these domains, compilation is not proof. Simulation, formal checking, hardware-in-the-loop tests, timing analysis, and traceability from requirement to verification evidence are essential.

\section{Method: Bounded Evidence Synthesis}
This paper uses a bounded evidence synthesis rather than a quantitative meta-analysis. The evidence base is heterogeneous: randomized trials, GitHub-scale datasets, repository-configuration studies, tool papers, issue-resolution surveys, hardware benchmarks, and process frameworks measure different outcomes. Pooling them into a single effect size would obscure rather than clarify the process problem.

Sources were selected from four streams:
\begin{enumerate}[leftmargin=*]
    \item agentic software engineering surveys and tool papers;
    \item empirical studies of developer productivity, agent-authored pull requests, and issue resolution;
    \item repository-configuration and execution-environment studies;
    \item hardware, firmware, and design-verification benchmarks.
\end{enumerate}

A source was included if it addressed at least one of the following: agentic code generation, issue resolution, repository-aware execution, productivity, configuration artifacts, permission or tool gating, hardware generation, hardware verification, lifecycle traceability, or verification evidence. Industry and platform sources were used only when they documented observable tool behavior or adoption trends; peer-reviewed and preprint research was used for empirical and conceptual claims.

\section{Problem Statement}
\subsection{The Limits of Conversational Development}
Conversational development is useful, but insufficient as an implementation substrate. It helps teams discover requirements, surface ambiguity, compare architectures, and identify risks. But long chat histories are not reliable engineering contracts:
\begin{itemize}[leftmargin=*]
    \item they contain superseded assumptions;
    \item constraints are often implicit;
    \item acceptance criteria are rarely executable;
    \item agents may overfit to recent turns;
    \item reviewers cannot easily audit which instruction governed a change.
\end{itemize}

Therefore, agentic work needs a conversation-to-contract gate: after discovery, the relevant intent must be converted into a reviewed execution brief.

\subsection{Verification Debt}
Agentic AI can increase the volume of code, tests, documentation, and patches. If output volume grows faster than verification capacity, teams accumulate verification debt: weak tests, hidden regressions, broad patches, unvalidated dependencies, undocumented behavior, and increased reviewer burden. The METR trial and maintenance-burden findings indicate that review and cleanup can erase perceived speedups \cite{metr_rct,maintenance_burden}. In hardware and embedded work, verification debt can become operational or physical risk.

\subsection{The Missing Bridge}
Current tools provide execution surfaces, sandboxing, repository instructions, test execution, and pull-request workflows. Existing engineering processes provide requirements discipline, verification logic, review, and release gates. The missing bridge is a lightweight framework that tells teams what input to provide to agents, how to structure agent execution, when to test, and which evidence is required before accepting generated artifacts.

\section{The Agentic Agile-V Framework}
\subsection{Overview}
\framework{} has two layers. The macro layer is Agile-V: an iterative lifecycle in which each increment remains traceable to requirements, design, implementation, verification, approval, and audit evidence. The micro layer is \scopev{}, the task-level loop used to run individual agentic tasks.

\begin{figure}[t]
\centering
\fbox{%
\begin{minipage}{0.88\linewidth}
\centering
\textbf{Agentic Agile-V}\vspace{0.35em}

\begin{tabular}{c}
Intent $\rightarrow$ Brief $\rightarrow$ Constraints \\
$\downarrow$ \\
Agentic execution via \scopev{} \\
$\downarrow$ \\
Tests, checks, simulation, review evidence \\
$\downarrow$ \\
Human approval and baseline update
\end{tabular}
\vspace{0.35em}

\parbox{0.95\linewidth}{\centering\itshape Conversation discovers intent. Structured artifacts control implementation. Evidence controls acceptance.}
\end{minipage}}
\caption{High-level \framework{} model.}
\label{fig:framework}
\end{figure}

This design avoids a false choice between agility and verification. Agile iteration provides speed; V-model reasoning provides traceability; agentic execution provides implementation capacity; verification gates decide acceptance.

\subsection{The \scopev{} Micro-Cycle}
Each agentic task passes through six steps:

\textbf{Specify.} Convert intent into a task brief with objective, scope, non-goals, affected modules, dependencies, acceptance criteria, and required evidence.

\textbf{Constrain.} Define boundaries: no public API change unless approved, no unrelated files, no new dependencies without justification, no broad refactor during a bug fix, explicit review for security-sensitive code, and mandatory preservation of hardware timing and safety constraints.

\textbf{Orchestrate.} Define how the agent should work: inspect first, summarize current design, propose a plan, implement small slices, run local checks, and produce a diff summary with residual risks.

\textbf{Prove.} Require evidence: unit tests, integration tests, regression tests, static analysis, type checks, linting, security scans, simulation, formal checks, hardware-in-the-loop results, or review checklists depending on risk.

\textbf{Evolve.} Feed validated learning back into repository instructions, templates, tests, and engineering baselines. Remove stale or harmful instructions.

\textbf{Verify.} Treat verification as recurring rather than final: before implementation, during patching, before merge, after deployment, and after field feedback.

\subsection{Agile-V as Lifecycle Backbone}
Agile-V was proposed to address a weakness of machine-speed AI-assisted engineering: lack of built-in task-level verification and regulatory traceability \cite{agilev}. Its case study demonstrates feasibility in a hardware-in-the-loop setting with independent test generation and audit artifacts. This paper generalizes the idea to agentic software, firmware, and hardware development by specifying inputs, task workflows, and acceptance gates.

The design principle is:
\begin{quote}
Agile-V controls the lifecycle; \scopev{} controls the agentic task.
\end{quote}

\section{Minimum Input Artifact Model}
Table \ref{tab:inputs} defines a minimum input package. The goal is not to overload agents with context, but to provide enough structured information to avoid guessing.

\begin{table*}[t]
\centering
\caption{Minimum input artifacts for agentic software, firmware, and hardware development.}
\label{tab:inputs}
\small
\begin{tabularx}{\textwidth}{p{0.17\textwidth}X X}
\toprule
\textbf{Artifact} & \textbf{Software Example} & \textbf{Firmware/Hardware Example} \\
\midrule
Intent and scope & User story, target behavior, non-goals, affected APIs & Function of module, hardware mode, peripheral role, integration boundary \\
Acceptance criteria & Expected inputs/outputs, UI behavior, error cases, performance target & Timing budget, protocol compliance, register behavior, waveform or HIL expectation \\
Architecture context & Modules, interfaces, dependencies, data model, service boundaries & Board revision, MCU/FPGA/SoC, memory map, clocks, buses, pinout, RTOS assumptions \\
Constraints & No API breakage, no new dependency, style and security rules & Voltage/power limits, timing constraints, interrupt rules, safety states, reset behavior \\
Execution context & Build commands, dependency setup, test commands, environment variables & Toolchain, simulator, synthesis tool, formal checker, test equipment, firmware flashing path \\
Evidence requirement & Unit/integration tests, lint, type checks, security scan, reviewer checklist & Simulation logs, formal properties, HIL logs, timing analysis, regression matrix \\
Risk class & Prototype, production, regulated, security-sensitive & Lab prototype, field firmware, safety-related, manufacturing release \\
\bottomrule
\end{tabularx}
\end{table*}

\section{Conversational Discovery vs. Structured Execution}
\subsection{When Conversation Helps}
Conversation is appropriate for early uncertainty: clarifying requirements, brainstorming architecture, identifying missing constraints, comparing test strategies, asking what could fail, and exploring alternative designs. In this phase the agent acts as a thinking partner. The desired output is not code first; it is a better problem statement.

\subsection{When Structure Is Mandatory}
Before implementation, the conversation must become a structured brief. This is mandatory when a task affects public APIs, safety, security, performance, hardware behavior, regulated workflows, customer-facing behavior, shared libraries, or persistent data.

The operational rule is:
\begin{quote}
Do not let an agent implement from a long chat. Let it implement from a reviewed brief.
\end{quote}

This rule is consistent with mixed evidence on context files. Relevant instructions can improve efficiency \cite{agentsmd_efficiency}; unnecessary or mismatched instructions can harm success \cite{agentsmd_eval}. The correct goal is not maximum context, but decision-relevant context.

\section{Task Workflows}
\subsection{Feature Development}
Feature work expands product intent into behavior. A recommended process is:
\begin{enumerate}[leftmargin=*]
    \item Write a feature brief: goal, non-goals, acceptance criteria.
    \item Identify affected modules, APIs, data structures, and tests.
    \item Ask the agent to inspect and summarize current design.
    \item Require a plan before edits.
    \item Implement the smallest useful slice.
    \item Add or update tests alongside implementation.
    \item Run targeted and regression checks.
    \item Produce a diff summary, evidence bundle, and residual-risk note.
    \item Require human review for architecture, security, maintainability, and edge cases.
\end{enumerate}

\subsection{Bug Fixing}
Bug fixing is causal diagnosis, not feature generation. The agent should not patch immediately. A recommended process is:
\begin{enumerate}[leftmargin=*]
    \item Capture observed and expected behavior.
    \item Provide reproduction steps, logs, environment, and version.
    \item Ask the agent for hypotheses and missing evidence.
    \item Localize likely files and call paths.
    \item Create a failing regression test where possible.
    \item Apply the minimal patch.
    \item Run regression and nearby tests.
    \item Explain why the fix works and what it does not address.
\end{enumerate}

Issue-resolution surveys emphasize that realistic maintenance requires long-horizon reasoning, iterative exploration, and feedback-driven decision-making beyond single-shot generation \cite{issue_resolution_survey,issue_resolution_frontier}.

\subsection{Testing and Review}
Testing must be inside the agent loop, not after it:
\begin{itemize}[leftmargin=*]
    \item \textbf{Before implementation:} identify expected tests, edge cases, and failure modes.
    \item \textbf{During implementation:} run targeted tests after small changes.
    \item \textbf{Before merge:} require CI, static analysis, type checks, security checks, and review.
    \item \textbf{After merge:} monitor logs, defects, performance, and user feedback.
\end{itemize}

Permission and action gating are part of the testing problem. A stress-test of Claude Code's auto mode found that permission-gate assumptions may fail under ambiguous state-changing scenarios, especially when equivalent effects can be achieved through file edits instead of shell commands \cite{permission_gate}. This supports explicit risk classification and human approval for state-changing or high-blast-radius actions.

\subsection{Hardware, Firmware, and Embedded Development}
Hardware-facing work requires stricter input and stricter evidence. The agent input package should include board/chip revision, datasheet excerpts, registers, pinout, clocks, protocols, memory map, toolchain, simulator, formal checker, test equipment, and rollback plan. Implementation should not be accepted without simulation, static checks, formal properties where applicable, and hardware-in-the-loop evidence for production-risk changes.

\section{Risk-Adaptive Acceptance Gates}
Not all tasks require the same rigor. Table \ref{tab:gates} defines four acceptance levels.

\begin{table*}[t]
\centering
\caption{Risk-adaptive acceptance gates for agent-generated artifacts.}
\label{tab:gates}
\small
\begin{tabularx}{\textwidth}{p{0.14\textwidth}X X X}
\toprule
\textbf{Risk Level} & \textbf{Typical Tasks} & \textbf{Required Evidence} & \textbf{Human Gate} \\
\midrule
R0: exploratory & Throwaway prototype, internal demo, draft script & Smoke test or manual run; no production credentials & Optional review \\
R1: routine & Documentation, small refactor, non-critical UI polish & Targeted tests, lint/type checks, diff summary & Normal code review \\
R2: production & Feature change, bug fix, public API, database migration & Regression tests, CI, static/security checks, rollback plan & Mandatory reviewer approval \\
R3: high assurance & Security, payments, medical, firmware, hardware, safety-related behavior & Traceable requirements, independent tests, simulation/formal/HIL as applicable, audit artifact & Explicit sign-off and release gate \\
\bottomrule
\end{tabularx}
\end{table*}

The acceptance rule is:
\begin{quote}
Agent output is not accepted because it is plausible; it is accepted because it satisfies evidence appropriate to its risk level.
\end{quote}

\subsection{Evidence Bundle}
For R2 and R3 tasks, the paper recommends a minimum evidence bundle:
\begin{itemize}[leftmargin=*]
    \item task brief and requirement identifiers;
    \item agent plan and affected files;
    \item executed commands and test results;
    \item diff summary and known residual risks;
    \item trace from acceptance criteria to tests;
    \item reviewer decision and follow-up actions;
    \item rollback or recovery path for production changes.
\end{itemize}

For hardware or firmware, the evidence bundle should include simulation logs, formal check results where applicable, HIL records, toolchain version, board revision, and timing or protocol evidence.

\section{Discussion}
\subsection{Implications for Teams}
Teams adopting agentic development should avoid both overconfidence and rejection. The practical steps are:
\begin{enumerate}[leftmargin=*]
    \item Maintain minimal repository instructions such as \texttt{AGENTS.md}, but keep them short, current, and testable.
    \item Use task templates for features, bugs, tests, and hardware work.
    \item Require agents to inspect, summarize, and plan before editing.
    \item Require tests or equivalent evidence before acceptance.
    \item Separate implementation and verification agents where risk is high.
    \item Track review load, defect escape, rework, and lead time rather than only code volume.
\end{enumerate}

\subsection{Implications for Tool Builders}
Tool builders should optimize not only for code generation but for evidence generation. Useful tool capabilities include structured brief editors, dependency setup capture, test discovery, traceability, risk classification, permission gates, sandboxed execution, review summaries, and exportable evidence bundles. The direction of the ecosystem, including OpenHands, GitHub agent workflows, Antigravity-style artifacts, and agent standards, is toward orchestration and observability rather than simple autocomplete \cite{openhands_sdk,github_agent_hq,antigravity,aaif}.

\subsection{Implications for Hardware and Embedded Engineering}
For embedded and hardware teams, the process should be even stricter. Agents can assist with drivers, register definitions, test harnesses, assertions, and documentation, but they must not bypass timing, protocol, safety, or HIL gates. Hardware benchmarks suggest that realistic system-level generation and verification remain difficult \cite{realbench,fixme,hardware_limits}.

\subsection{Threats to Validity}
This paper is a synthesis and process proposal, not a new benchmark. The field is moving quickly, and point estimates from 2024 to 2026 may change as models and tools evolve. Productivity studies vary by task type, developer experience, codebase maturity, tool generation, and organizational culture. Hardware benchmarks differ in design complexity and verification rigor. The framework should therefore be validated empirically in future studies across multiple teams, repositories, tools, and hardware domains.

\section{Practical Templates}
\subsection{Feature Brief Template}
A feature brief should include: objective, user-visible behavior, non-goals, affected modules, interface contracts, migration needs, compatibility constraints, security considerations, acceptance criteria, tests to add or update, and rollback path.

\subsection{Bug Brief Template}
A bug brief should include: observed behavior, expected behavior, reproduction steps, input data, logs/traces, environment, affected version, suspected area, failing test if available, constraints on fix scope, and regression-test requirement.

\subsection{Hardware/Firmware Brief Template}
A hardware or firmware brief should include: board revision, chip or FPGA variant, datasheet excerpts, register map, pinout, clock tree, bus/protocol rules, memory map, RTOS/bare-metal assumptions, timing and power constraints, safety states, toolchain, simulator, HIL setup, and acceptance evidence.

\section{Research Agenda}
Future work should evaluate \framework{} empirically. Key questions include:
\begin{enumerate}[leftmargin=*]
    \item Do structured execution briefs improve agent success compared with conversational prompts?
    \item What is the minimum useful content of repository instructions?
    \item Does independent agentic test generation reduce defect escape?
    \item Which task classes benefit from Agile-V/\scopev{} and which are slowed by overhead?
    \item Can evidence bundles reduce verification debt without eliminating productivity gains?
    \item How should hardware and firmware evidence quality be measured?
\end{enumerate}

\section{Conclusion}
Agentic AI is changing software and hardware development, but it does not make engineering process obsolete. The evidence rejects both extremes: agentic coding is not merely a toy, but neither is it a universal productivity multiplier. Its value depends on context, constraints, execution environments, verification gates, and human oversight.

This paper proposed \framework{}, combining Agile-V lifecycle discipline with the \scopev{} task loop. The central message is that conversation is good for discovering intent, but structured artifacts are required for implementation. Code, tests, documentation, firmware, and hardware designs should be accepted only when they produce evidence appropriate to their risk level. The future of agentic engineering is not vibe coding at scale. It is verified engineering with agents inside the loop.

\end{document}